\documentclass[]{aastex631}

\usepackage{mathptmx}
\usepackage{amsmath}
\usepackage{threeparttable}
\usepackage{xcolor}
\usepackage{booktabs}
\usepackage{graphicx}
\usepackage{rotating}
\begin{document}

\title{Binary mass transfer fossil in the ancient GD-1 stellar stream}

\author{Guochao Yang}
\altaffiliation{Contact e-mail: gcyang@nao.cas.cn}
\affiliation{School of Physics and Astronomy, China West Normal University, Nanchong 637009, People's Republic of China}
\affiliation{National Astronomical Observatories, Chinese Academy of Sciences, Beijing 100012, People's Republic of China}
\author{Sofya Alexeeva}
\altaffiliation{Contact e-mail: alexeeva@nao.cas.cn}
\affiliation{National Astronomical Observatories, Chinese Academy of Sciences, Beijing 100012, People's Republic of China}
\author{Jingkun Zhao}
\altaffiliation{Contact e-mail: zjk@nao.cas.cn}
\affiliation{National Astronomical Observatories, Chinese Academy of Sciences, Beijing 100012, People's Republic of China}
\author{Wako Aoki}
\affiliation{National Astronomical Observatory of Japan, 2-21-1 Osawa, Mitaka, Tokyo 181-8588, Japan}
\affiliation{Astronomical Science Program, The Graduate University for Advanced Studies, SOKENDAI, 2-21-1 Osawa, Mitaka, Tokyo 181-8588, Japan}
\author{Tatyana Sitnova}
\affiliation{Institute of Astronomy, Russian Academy of Sciences, Moscow, 119017, Russia}
\author{Xianhao Ye}
\affiliation{National Astronomical Observatories, Chinese Academy of Sciences, Beijing 100012, People's Republic of China}
\affiliation{Institute of Astronomy and Physics, Inner Mongolia University, Hohhot 010021, People’s Republic of China}
\author{Nian Liu}
\affiliation{School of Physics and Astronomy, China West Normal University, Nanchong 637009, People's Republic of China}
\author{Qianfan Xing}
\affiliation{National Astronomical Observatories, Chinese Academy of Sciences, Beijing 100012, People's Republic of China}
\author{Gang Zhao}
\affiliation{National Astronomical Observatories, Chinese Academy of Sciences, Beijing 100012, People's Republic of China}
\affiliation{Institute of Space Sciences, School of Space Science and Technology, Shandong University, Weihai 264209,
People's Republic of China}
\begin{abstract}

We report the discovery of J0930+2902, the first carbon-enhanced metal-poor star with s-process enrichment (CEMP-s)
identified in the GD-1 stellar stream, which is widely believed to originate from the tidal disruption of an old,
metal-poor globular cluster. Using high-resolution Subaru/HDS spectroscopy, we present a comprehensive chemical
abundance analysis of this star. Stellar atmospheric parameters were re-derived using Gaia photometry and parallaxes,
combined with an iterative non-local thermodynamic equilibrium (NLTE) analysis of Fe lines. We determine abundances for 22
elements, including NLTE analyses of C, Na, Mg, Ca, Ti, Cr, Mn, Co, Sr, Y, Zr, Ba, and Eu. J0930+2902 exhibits
pronounced enhancements of C, N, Na, and s-process elements, together with a low $^{12}$C/$^{13}$C ratio of 7.4,
consistent with enrichment through binary mass transfer from a former asymptotic giant branch (AGB) companion.
Large radial-velocity variations and a significant near-ultraviolet excess independently indicate the presence
of a white dwarf companion, providing strong support for this scenario. Comparisons with FRUITY AGB nucleosynthesis
models show that the residual neutron-capture abundance pattern is best reproduced by a low-mass AGB
companion with an initial mass of approximately 3.0~$M_{\odot}$. Its enhancements of C, N, Na, and s-process
elements make it chemically distinct from the other confirmed GD-1 members. These results identify J0930+2902 as a
fossil record of binary mass transfer preserved in the ancient GD-1 stellar stream, demonstrating that binary
interaction can leave long-lived chemical signatures in accreted stellar populations.

\end{abstract}


\section{Introduction}

The GD-1 stellar stream stands out as one of the thinnest and dynamically coldest structures in the Milky Way halo,
spanning over 80$^\circ$ with a remarkable length-to-width ratio exceeding 100:1 \citep{Koposov2010}. Since its
discovery in Sloan Digital Sky Survey (SDSS) data \citep{York2000,Grillmair2006}, GD-1 has served as a pristine
laboratory for mapping the Galactic gravitational potential and probing dark matter substructure
\citep{Malhan2019,Bonaca2019,Bonaca2020,Carlberg2026,Jarvis2026}. Photometric, spectroscopic, and dynamical
studies indicate that GD-1 is a metal-poor and chemically homogeneous stellar population originating from a low-mass
globular cluster (GC), with an initial stellar mass of order $10^4~M_\odot$ \citep{Willett2009,Li2018,Boer2020,Tavangar2025}.
The progenitor has undergone substantial tidal disruption and is now largely or completely dissolved. N-body modeling
by \citet{Webb2019} found that the observed GD-1 stream is consistent with a dynamical age of approximately 2--3~Gyr.
The same study identified two viable disruption scenarios, with the progenitor either dissolving approximately 0.5~Gyr
ago or more than 2.5~Gyr ago. These timescales characterize the dynamical evolution of the stream and do not uniquely
establish when its progenitor was accreted by the Milky Way, leaving its detailed accretion and disruption history
uncertain. GD-1 therefore provides an excellent setting for studying the chemical and binary evolution of stars
originating in a disrupted GC.

Recently, in our high-resolution spectroscopic study \citep[][hereafter Paper~I]{Zhao2025} of seven GD-1 member stars
observed with the High Dispersion Spectrograph (HDS) mounted on the Subaru Telescope \citep{Noguchi2002}, we found that
six stars exhibit chemically homogeneous abundance patterns, with a mean metallicity of [Fe/H] $\sim -2.38$ and
enhanced Eu abundances. These chemical properties further support the hypothesis that GD-1 originated from a disrupted
low-mass GC. Among these stars, one object, LAMOST J093041.07+290158.7 (hereafter J0930+2902), shows remarkable
chemical peculiarities, including strong carbon enhancement and unusually high abundances of neutron-capture elements
such as Y and Ba. J0930+2902 was previously identified as a carbon star by \citet{Ji2016} based on low-resolution
spectroscopy from the Large Sky Area Multi-Object Fiber Spectroscopic Telescope (LAMOST; \citealt{Cui2012,Zhao2012,Luo2012}),
but its detailed chemical abundance pattern and nucleosynthetic origin have not yet been systematically investigated.

Elemental abundances in stars preserve the signatures of the nucleosynthetic processes responsible for their
production \citep{Sneden2008,Arcones2023}. While light elements, $\alpha$-elements, and iron-group elements are
synthesized through stellar fusion reactions, elements heavier than the iron-group are primarily produced through
neutron-capture processes \citep{Burbidge1957}. These include the slow (s-) and rapid (r-) neutron-capture processes.
The main s-process operates in low- and intermediate-mass asymptotic giant branch (AGB) stars
\citep{Busso1999,Herwig2005}, whereas the r-process operates at much higher neutron densities. Neutron star mergers
are now observationally established as a site of main r-process production, while various types of supernovae
have been proposed as possible sites \citep{Winteler2012,Thielemann2017,Watson2019}. At solar metallicity, Ba, La,
and Ce are predominantly produced by the s-process, whereas Eu is mainly produced by the r-process.
However, the relative contributions of the two processes can differ substantially at low metallicity \citep{Arcones2023}.

The chemical characteristics of metal-poor stars ([Fe/H] $< -1.0$) provide insights into the chemical evolution of
the early Galaxy \citep{Beers2005}. Observations indicate that a substantial fraction of metal-poor stars in the
Galactic halo exhibits large overabundances of carbon ([C/Fe] $>$ 0.7], \citealt{Aoki2007}). These stars are referred
to as carbon-enhanced metal-poor (CEMP) stars. According to the Preston/Shectman
\cite[HK,][]{Beers1985,Beers1992,Beers1999} and Hamburg/ESO objective-prism \cite[HES,][]{Wisotzki2000} surveys,
about 20\% of halo stars with [Fe/H] $< -2.0$ show [C/Fe] $>$ 1.0 \citep{Placco2014}. CEMP stars are
commonly classified into four subcategories based on their abundances of neutron-capture elements: CEMP-no stars, which
show no enhancements in either s-process or r-process elements ([Ba/Fe] $<$ 0.0); CEMP-r stars, which exhibit
overabundances in r-process elements ([Eu/Fe] $>$ 1.0); CEMP-s stars, which display overabundances in s-process
elements ([Ba/Fe] $>$ 1.0 and [Ba/Eu] $>$ 0.5); and CEMP-r/s stars, which show overabundances in both groups
(0.0 $<$ [Ba/Eu] $<$ 0.5, \citealt{Beers2005,Frebel2018,Goswami2021}). Radial-velocity (RV) variations in CEMP-s
stars suggest that they are members of binary systems, with their s-process element enhancements resulting
from mass transfer from a companion that produced carbon and s-process elements during its AGB phase
\citep{Lucatello2005,Starkenburg2014,Goswami2021}.

Although our Paper~I reported local thermodynamic equilibrium (LTE) abundances for several atomic species
in J0930+2902, several critical abundance
diagnostics remained unexplored. These include the carbon abundance inferred from the \ion{C}{1} lines and CH
and C$_2$ molecular bands, the nitrogen abundance inferred from the CN molecular bands, and the abundances of
key neutron-capture elements such as Sr, Zr, La, Ce, Pr, Nd, Sm, and Eu.
Together with the isotopic ratio $^{12}$C/$^{13}$C, which provides a sensitive tracer of nucleosynthesis and
internal mixing, these quantities are crucial for establishing the detailed abundance pattern of the star and
clarifying its nucleosynthetic origin \citep{Abia2017}. A comprehensive abundance analysis is therefore
required to determine whether J0930+2902 is a genuine CEMP-s star formed via mass transfer from a former AGB companion.

As the first s-process-enhanced star identified in the GD-1 stellar stream, J0930+2902 offers a unique opportunity
to probe binary evolution and neutron-capture nucleosynthesis in an accreted, metal-poor stellar stream environment.
We present an updated abundance analysis of J0930+2902 based on high-resolution Subaru/HDS spectra. In Section~2,
we describe the methodology used for the abundance determination. Section~3 presents and discusses the resulting
abundances. Section~4 examines the binary nature of J0930+2902, and Section~5 estimates the companion mass.
Section~6 discusses the implications of J0930+2902 for the GD-1 progenitor. Section~7 summarizes the main conclusions.

\section{Abundance Determination}

As described in Paper~I, our program stars were selected from candidate GD-1 members
using LAMOST \citep{Zhao2012} and SEGUE \citep{Yanny2009} spectroscopy combined with
Gaia~DR3 astrometry \citep{Gaia2023}, following the membership selection method of
\citet{Li2018}. J0930+2902 is one of the seven GD-1 members studied in Paper~I and is
revisited here with a detailed analysis of its chemical abundances and binary nature.
High-resolution spectra ($R \sim 36{,}000$) of J0930+2902 were obtained with Subaru/HDS,
covering the wavelength range 4030--6800~\AA\ \citep{Noguchi2002}. Details of the
spectroscopic observations were presented in Paper I.

\subsection{Stellar atmospheric parameters}

The stellar atmospheric parameters were derived through an iterative analysis
based on Gaia photometry, parallax, and non-local thermodynamic equilibrium (NLTE) spectroscopy.
The effective temperature
($T_{\rm eff}$) was initially estimated from the dereddened Gaia BP$-$G, G$-$RP, and
BP$-$RP colors using the calibration of \citet{Mucciarelli2021}. The Gaia magnitudes
are G = 14.2714~mag, BP = 14.6656~mag, and RP = 13.7065~mag. We adopted a reddening
of E(B-V) = 0.016~mag from \citet{Schlafly2011}, with the corresponding correction
following \citet{Casagrande2018}. The three color indices yielded consistent initial
temperature estimates, giving $T_{\rm eff} = 5166$~K.

The initial surface gravity ($\log g$) was estimated from the Gaia parallax,
corrected for the zero-point offset following \citet{Lindegren2021}. The measured parallax
is $\pi = 0.0586 \pm 0.0265$~mas. The distance was derived from the maximum of the posterior
probability distribution function \citep{Bailer2015}. Together with the initial
$T_{\rm eff}$ and the bolometric correction from \citet{Casagrande2018}, it was used to
calculate an initial surface gravity of $\log g = 2.06$. Because the parallax uncertainty
exceeded 20\%, we treated the parallax-based $\log g$ as an initial estimate
and subsequently refined it using 12~Gyr isochrones with [Fe/H] = $-2.5$ and
[$\alpha$/Fe] = 0.4 \citep{Dotter2008}.

\begin{figure*}
 \begin{center}
 \includegraphics[scale=0.4]{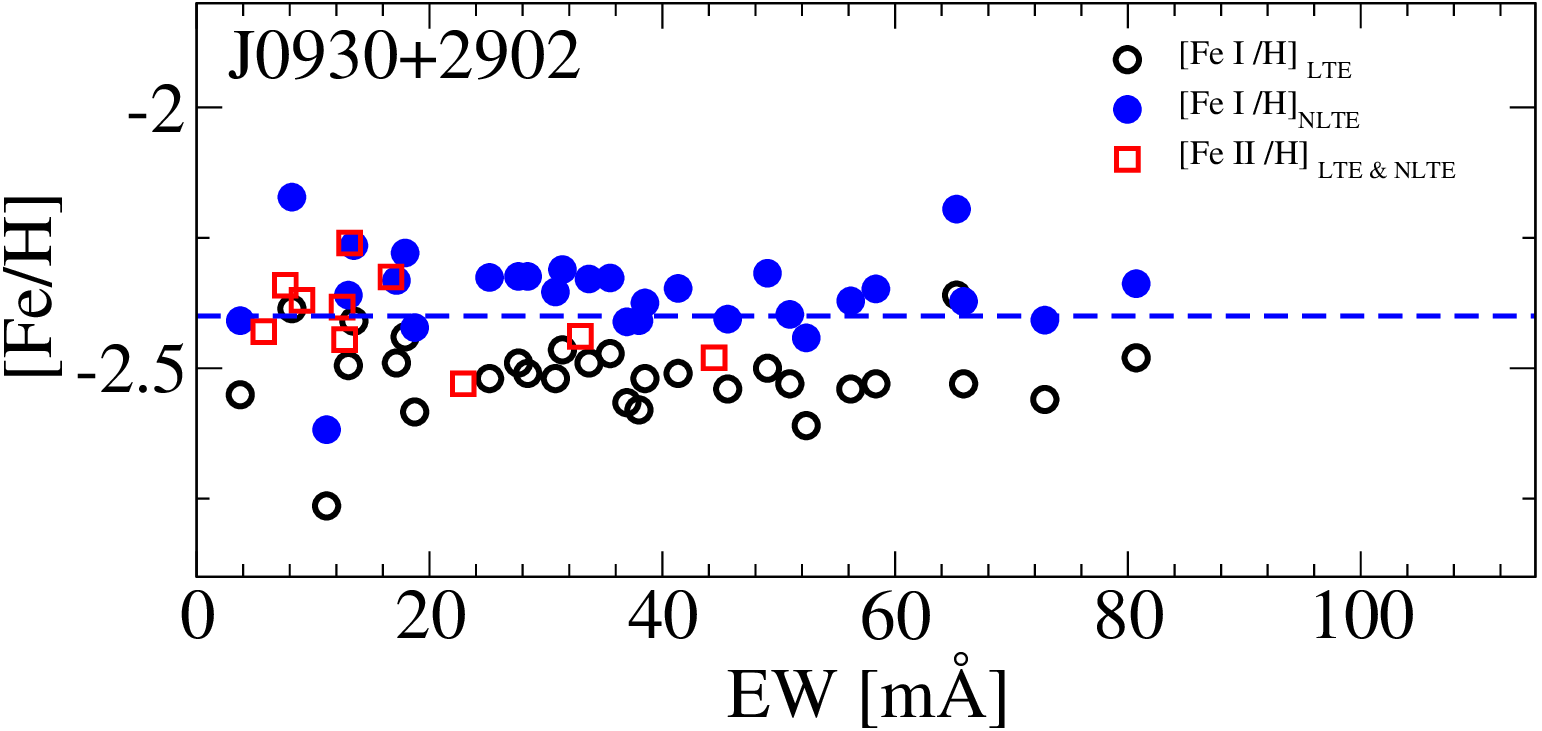}
 \caption{[Fe/H] derived from \ion{Fe}{1} and \ion{Fe}{2} lines as a function of
 equivalent width (EW). The black open circles and blue filled circles represent the LTE and
 NLTE abundances derived from \ion{Fe}{1} lines, respectively. The red squares represent the
 \ion{Fe}{2} abundances derived under both LTE and NLTE assumptions, which are effectively
 identical and are therefore shown with the same symbol. The dashed line indicates the adopted [Fe/H].}
 \label{FeH}
 \end{center}
 \end{figure*}

We adopted an initial metallicity of [Fe/H] = $-2.36$ from Paper~I.
NLTE corrections for \ion{Fe}{1} were obtained from the INASAN database \citep{Mashonkina2023}.
For \ion{Fe}{2} lines, departures from LTE are expected to be negligible \citep{Mashonkina2016}.
We determined the final stellar parameters iteratively using the NLTE abundances of \ion{Fe}{1}
and \ion{Fe}{2}. The effective temperature was refined by imposing excitation equilibrium, i.e.,
by requiring the iron abundances to show no dependence on excitation potential (EP). The surface
gravity was then iteratively refined by imposing ionization equilibrium, i.e., by requiring
consistency between the abundances derived from \ion{Fe}{1} and \ion{Fe}{2} lines.
The microturbulent velocity ($\xi_{\rm t}$) was determined by requiring no dependence of
the iron abundances on equivalent width (EW). As shown in Figure~\ref{FeH}, we obtained
[\ion{Fe}{1}/H] = $-2.35 \pm 0.08$ from 29 \ion{Fe}{1} lines and
[\ion{Fe}{2}/H] = $-2.40 \pm 0.08$ from 10 \ion{Fe}{2} lines.
The final stellar parameters derived for J0930+2902 are $T_{\rm eff} = 5215$~K, $\log g = 2.04$,
[Fe/H] = $-2.40$, and $\xi_{\rm t}$ = 1.70~km\,s$^{-1}$.

\subsection{NLTE calculations}

Abundances of C, Na, Mg, Ca, Ti, Sr, Y, Zr, Ba, and Eu were determined in this study using
NLTE line-formation calculations, with the model atoms and methods developed by \citet{Mashonkina1999,
Mashonkina2001,Mashonkina2007,Velichko2010,Alexeeva2014,Alexeeva2015,Sitnova2016,Alexeeva2018,
Alexeeva2023,Mashonkina2023}. For Cr, Mn, and Co, we adopted the NLTE abundance corrections from the
MPIA online NLTE abundance correction tool\footnote{\url{https://nlte.mpia.de/gui-siuAC_secE.php}},
based on the calculations of \citet{BergemannCescutti2010}, \citet{Bergemann2010}, and \citet{Bergemann2019}.

For \ion{Sr}{2} and \ion{Y}{1}--\ion{Y}{2}, inelastic collisions with neutral hydrogen
particles were treated using rate coefficients derived from the quantum-mechanical calculations
of \citet{Yakovleva2022} and \citet{Wang2024}, respectively. For \ion{Zr}{2}, inelastic
collisions with neutral hydrogen were treated using the classical Drawin formula \citep{Drawin1968},
as implemented by \citet{Steenbock1984}. We adopted a scaling factor of $S_{\rm H} = 1$,
as recommended by \citet{Velichko2010}.

We derived Sr abundances assuming a pure $^{88}$Sr isotopic composition. Hyperfine structure (HFS)
was neglected for the \ion{Y}{2} lines because the HFS splitting of the sole stable isotope, $^{89}$Y,
is less than 1~m\AA\ and has a negligible impact on the derived abundances at the resolution of our observations
\citep{Hannaford1982,Villemoes1992}. For Zr, Nd, and Eu, we adopted the r-process isotopic mixtures
from \citet{Prantzos2020}, with abundance ratios of $^{90}$Zr:$^{91}$Zr:$^{92}$Zr:$^{94}$Zr:$^{96}$Zr = 55:3:24:7:11, $^{143}$Nd:$^{144}$Nd:$^{145}$Nd:$^{146}$Nd:$^{148}$Nd:$^{150}$Nd = 19:26:16:13:12:14, and
$^{151}$Eu:$^{153}$Eu = 48:52, respectively. For Ba and Sm, we adopted the protosolar isotopic mixtures
from \citet{Lodders2021}, with approximate isotopic abundance ratios of
$^{134}$Ba:$^{135}$Ba:$^{136}$Ba:$^{137}$Ba:$^{138}$Ba = 2:7:8:11:72 and $^{144}$Sm:$^{147}$Sm:$^{148}$Sm:$^{149}$Sm:$^{150}$Sm:$^{152}$Sm:$^{154}$Sm = 3:15:11:14:7:27:23, respectively.
Both isotopic shifts (IS) and HFS were included in the synthesis of the \ion{Ba}{2} subordinate lines, with HFS
included for the odd isotopes, $^{135}$Ba and $^{137}$Ba \citep{Wendt1984}.

The radiative transfer and statistical equilibrium equations were simultaneously solved
using the \textsc{DETAIL} code \citep{Butler1984} with the updated opacity package introduced by \citet{Mashonkina2011}.
The resulting departure coefficients, $b_{\rm i} = n_{\rm NLTE}/n_{\rm LTE}$, were applied to compute
theoretical NLTE line profiles using the \textsc{synthV\_NLTE} code \citep{Tsymbal2019}.
Here, $n_{\rm NLTE}$ and $n_{\rm LTE}$ are the statistical equilibrium and thermal (Saha--Boltzmann) number
densities, respectively.

\section{Results}
\subsection{Chemical composition}

The chemical abundances of J0930+2902 were derived mainly through spectral synthesis with
BINMAG \citep{Kochukhov2010,Kochukhov2018}. The Si, Sc, Pr, and Sm abundances were additionally derived using
iSpec \citep{BlancoCuaresma2014}, adopting the same stellar atmospheric parameters as those used in the main
abundance analysis. We adopted MARCS model atmospheres \citep{Gustafsson2008} and atomic data mainly from
the latest version of the Vienna Atomic Line Database (VALD; \citealt{Ryabchikova2015,Pakhomov2019}),
supplemented by published atomic data for specific lines, and included isotopic and HFS
components where relevant. NLTE abundances were determined for C, Na, Mg, Ca, Ti,
Cr, Mn, Co, Sr, Y, Zr, Ba, and Eu, while LTE abundances were adopted for the remaining species.
Individual lines used in this study, along with their EPs,
oscillator strengths ($\log gf$), references, and derived abundances are presented in Table~\ref{abn1}.
For \ion{Fe}{2}, we adopted the $\log gf$ values from \citet{RaassenUylings1998}, with a correction
of +0.11~dex following the recommendation of \citet{Grevesse1999}. Throughout this work, abundances are
reported on the scale where $A({\rm H}) = 12$, and abundance ratios [X/Fe] are given in Table~\ref{abn2}.
Solar reference abundances were adopted from \citet{Asplund2009}.

\begin{figure*}
   \begin{minipage}{170mm}
 \begin{center}
 \parbox{0.32\linewidth}{\includegraphics[scale=0.19]{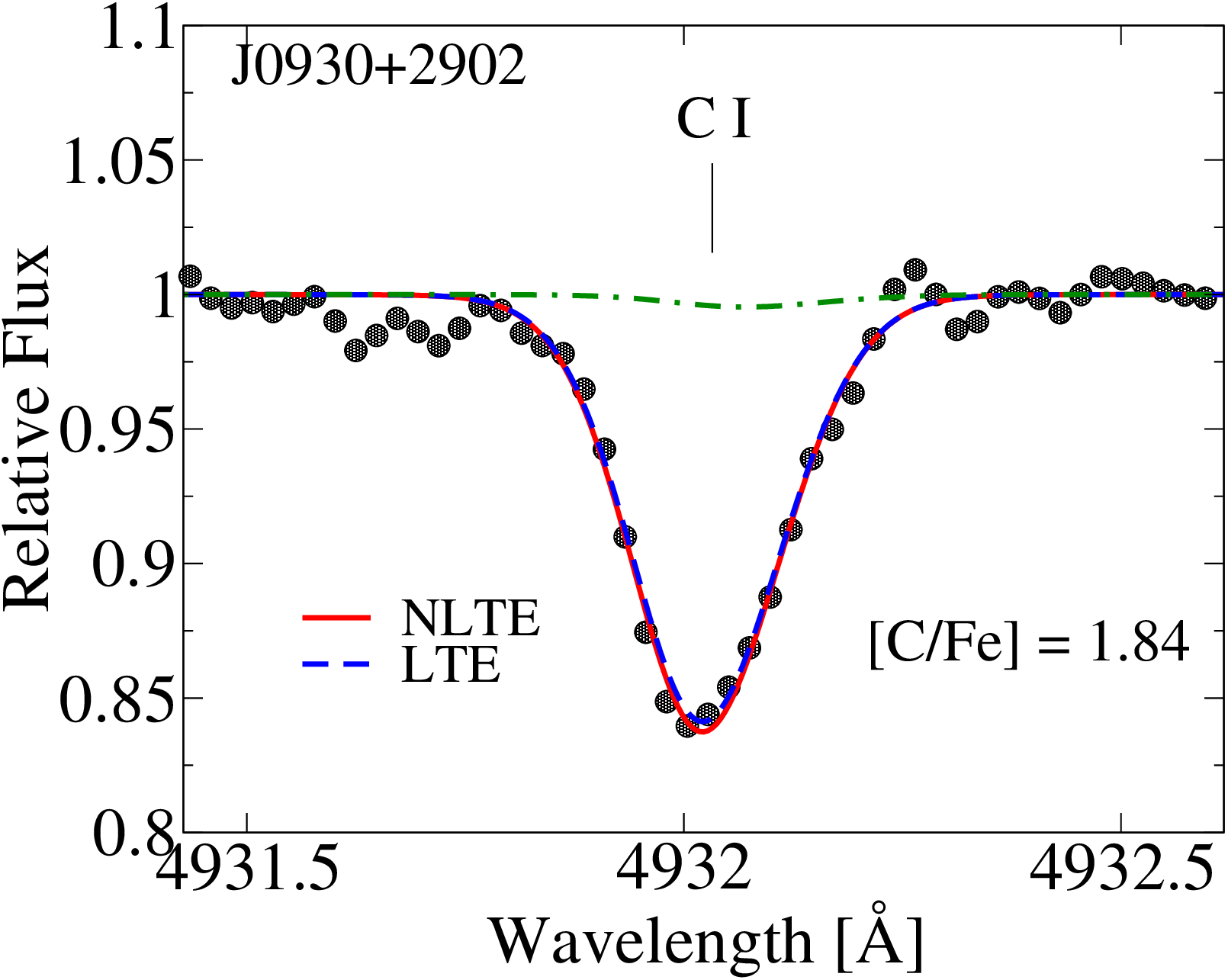}\\
  \centering}
   \parbox{0.32\linewidth}{\includegraphics[scale=0.19]{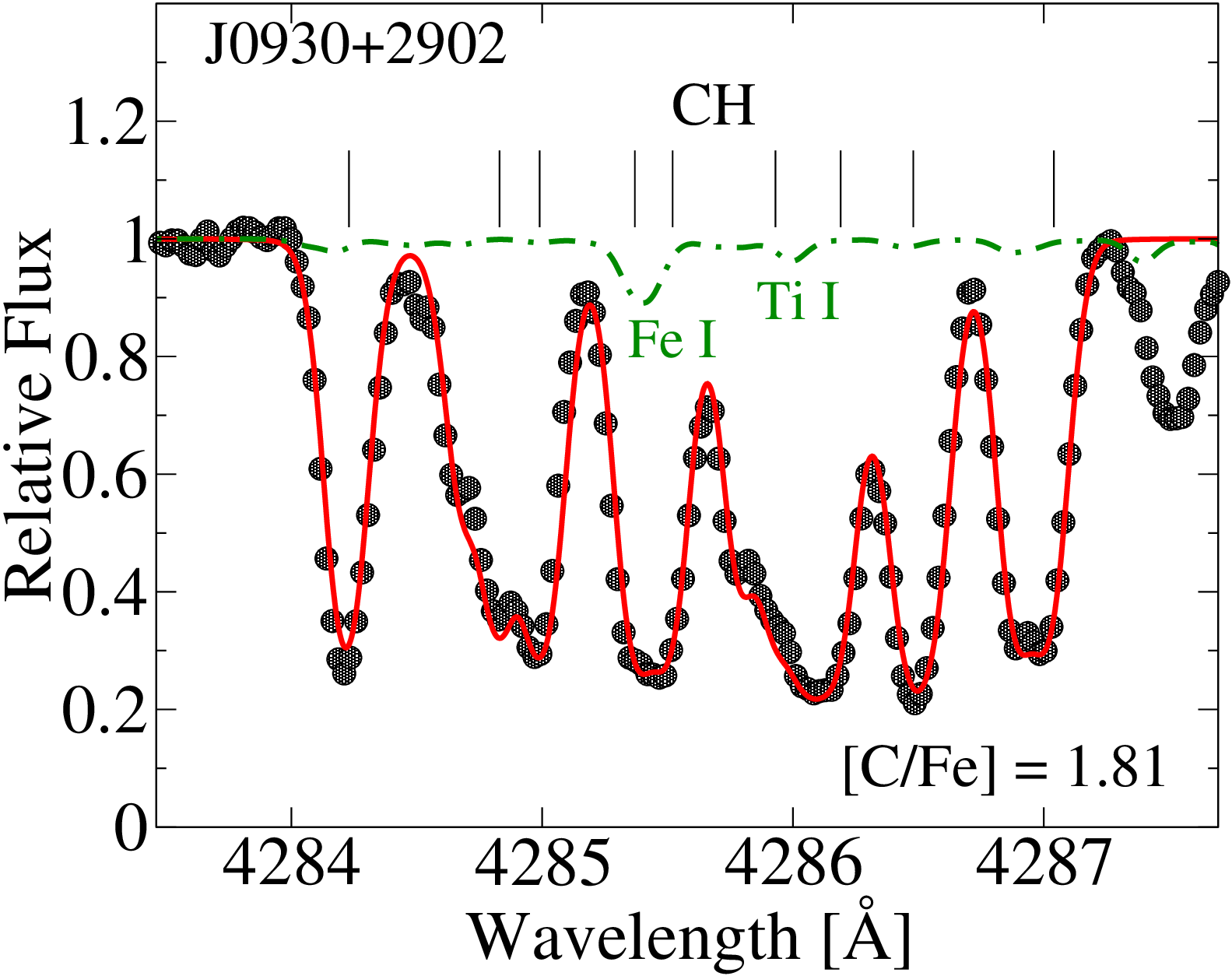}\\
  \centering}
 \parbox{0.32\linewidth}{\includegraphics[scale=0.19]{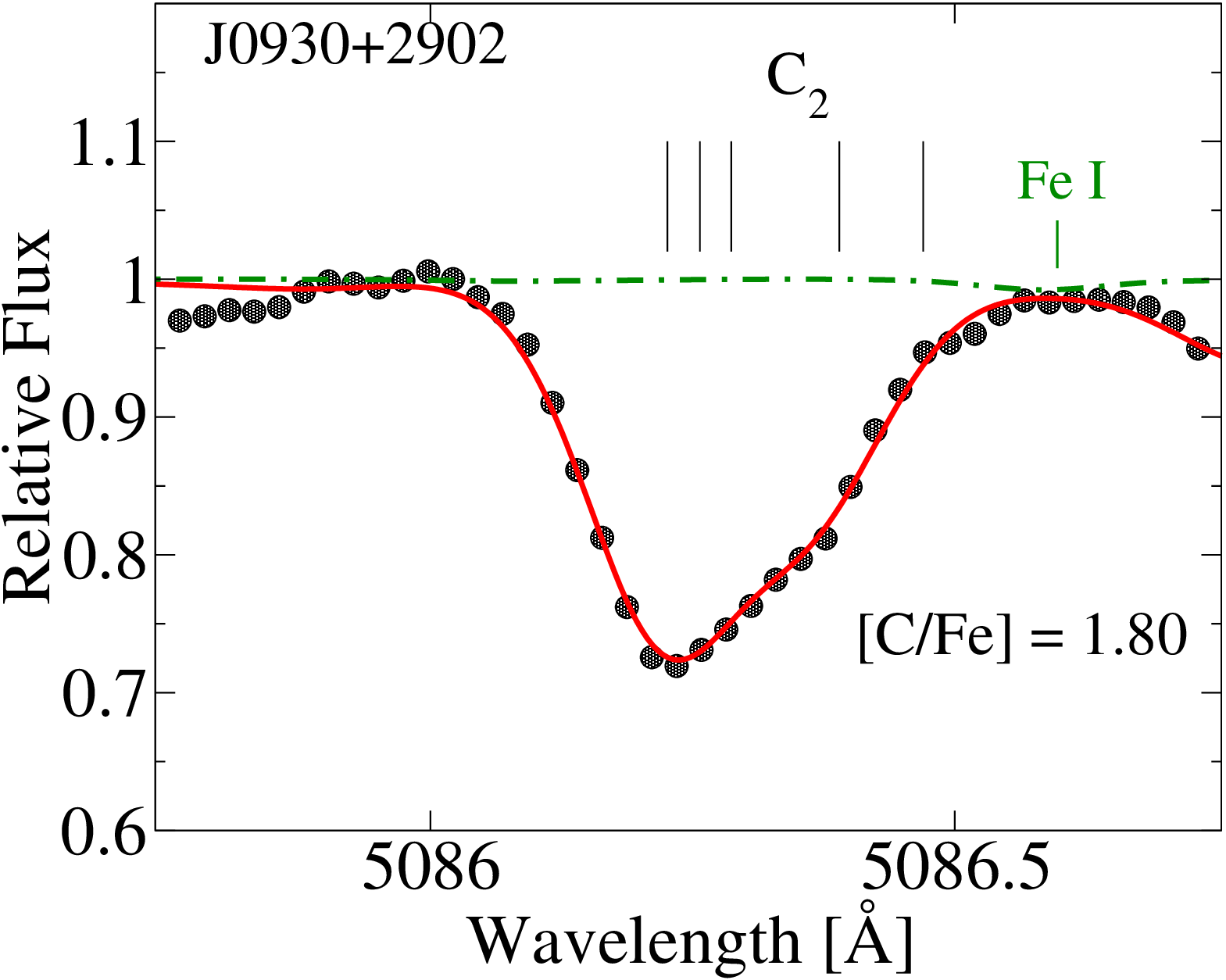}\\
 \centering}
 \hspace{1\linewidth}
 \hfill
 \\[0ex]
 \caption{Best fits of selected atomic and molecular lines in J0930+2902.
 The observed spectra are shown by bold dots. The symbol sizes indicate the typical
 uncertainties of the observed data points. The red solid curves represent the
 best-fit NLTE synthetic spectra, while the blue dashed curves show the LTE synthetic
 spectra calculated assuming the same abundances as in the NLTE analysis. The green
 dash-dotted curves represent synthetic spectra calculated with the abundance of the
 element under investigation set to zero.}
 \label{Carbon}
 \end{center}
 \end{minipage}
 \end{figure*}

{\bf Carbon.}
Carbon abundances were derived independently from the \ion{C}{1} lines at 4932 and 5052~\AA, the CH
molecular features, and the C$_2$ bands. We adopted [O/Fe] = 0.50, consistent with
the oxygen abundance plateau found for metal-poor stars by \citet{SitnovaTM2016}. The resulting abundances
are in excellent agreement, yielding [C/Fe] = $1.84 \pm 0.04$ from \ion{C}{1}, [C/Fe] = $1.80 \pm 0.02$
from C$_2$, and [C/Fe] = $1.80 \pm 0.02$ from CH. Representative fits are shown in Figure~\ref{Carbon}.

\begin{figure*}
 \begin{center}
 \includegraphics[scale=0.4]{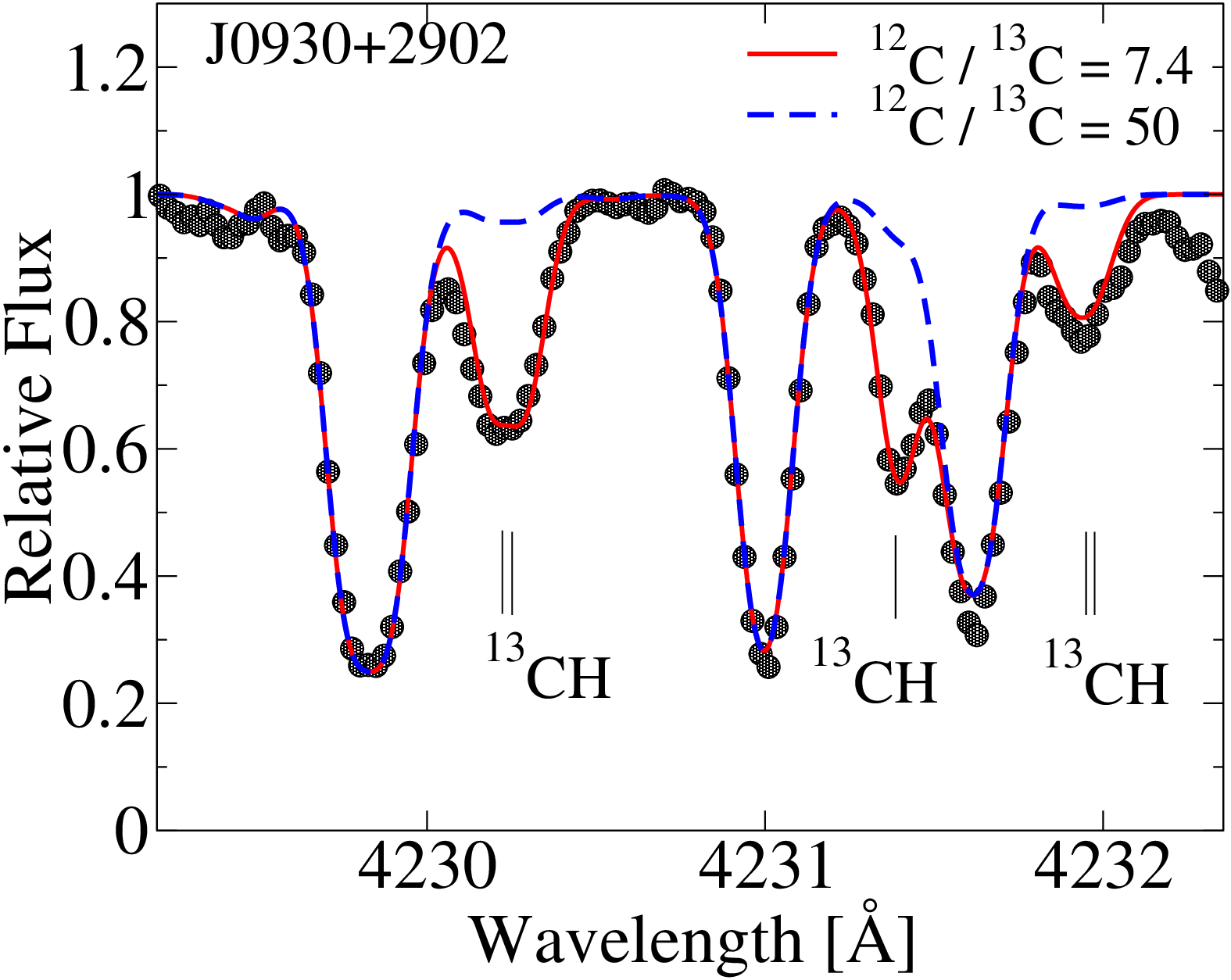}
 \caption{The narrow spectral region of J0930+2902 around the G-band is shown, together with two
 synthetic spectra in different colors corresponding to different carbon isotopic ratios.
 The symbol sizes indicate the typical uncertainties of the corresponding data points.}
 \label{13C}
 \end{center}
 \end{figure*}

{\bf $^{12}$C/$^{13}$C ratio.}
The carbon isotopic ratio was estimated from spectrum synthesis of the CH $A^2\Delta$--$X^2\Pi$ (G-band)
features between 4150 and 4350~\AA. We derived $^{12}$C/$^{13}$C = 7.4. Figure~\ref{13C} shows the spectral
region used in the analysis.

{\bf Nitrogen and Sodium.}
Nitrogen abundances were derived from CN molecular features while fixing the carbon abundance at the
value determined above. We obtained [N/Fe] = $2.06 \pm 0.05$, indicating strong nitrogen enhancement.
From four \ion{Na}{1} lines at $\lambda\lambda$5682, 5688, 6154, and 6160~\AA, we derived an NLTE
abundance ratio of [Na/Fe] = $1.54 \pm 0.10$, revealing a substantial overabundance of sodium relative
to iron.

{\bf Upper limits of O and Pb abundances.}
For the light element O and the heavy element Pb, we derived upper limits of [O/Fe] $\leq$ 0.51 from
the \ion{O}{1} line at $\lambda$6300~\AA\ and [Pb/Fe] $\leq$ 1.74 from the \ion{Pb}{1} line at
$\lambda$4057~\AA, respectively.

{\bf $\alpha$- and iron-group elements.}
We determined NLTE abundances for the $\alpha$-elements Mg, Ca, and Ti, and the iron-group elements Cr
and Co, while LTE abundances were derived for Si, Sc, Mn, and Ni. HFS was taken into
account for Sc, Mn and Co. The derived [X/Fe] ratios range from $-0.11$ to 0.68.

{\bf Neutron-capture elements.}
From spectral synthesis of the \ion{Sr}{2} $\lambda\lambda$4077 and 4215 lines, we derived
[Sr/Fe] = $1.45 \pm 0.23$. Analysis of the \ion{Y}{2} $\lambda\lambda$4883, 5087, and 5205 lines yielded
[Y/Fe] = $1.54 \pm 0.09$, while the \ion{Zr}{2} $\lambda\lambda$4442, 4629, and 5112 lines yielded
[Zr/Fe] = 2.02. The large enhancements of Sr, Y, and Zr indicate strong enrichment in first-peak
neutron-capture elements.

\begin{figure*}
   \begin{minipage}{170mm}
 \begin{center}
 \parbox{0.32\linewidth}{\includegraphics[scale=0.2]{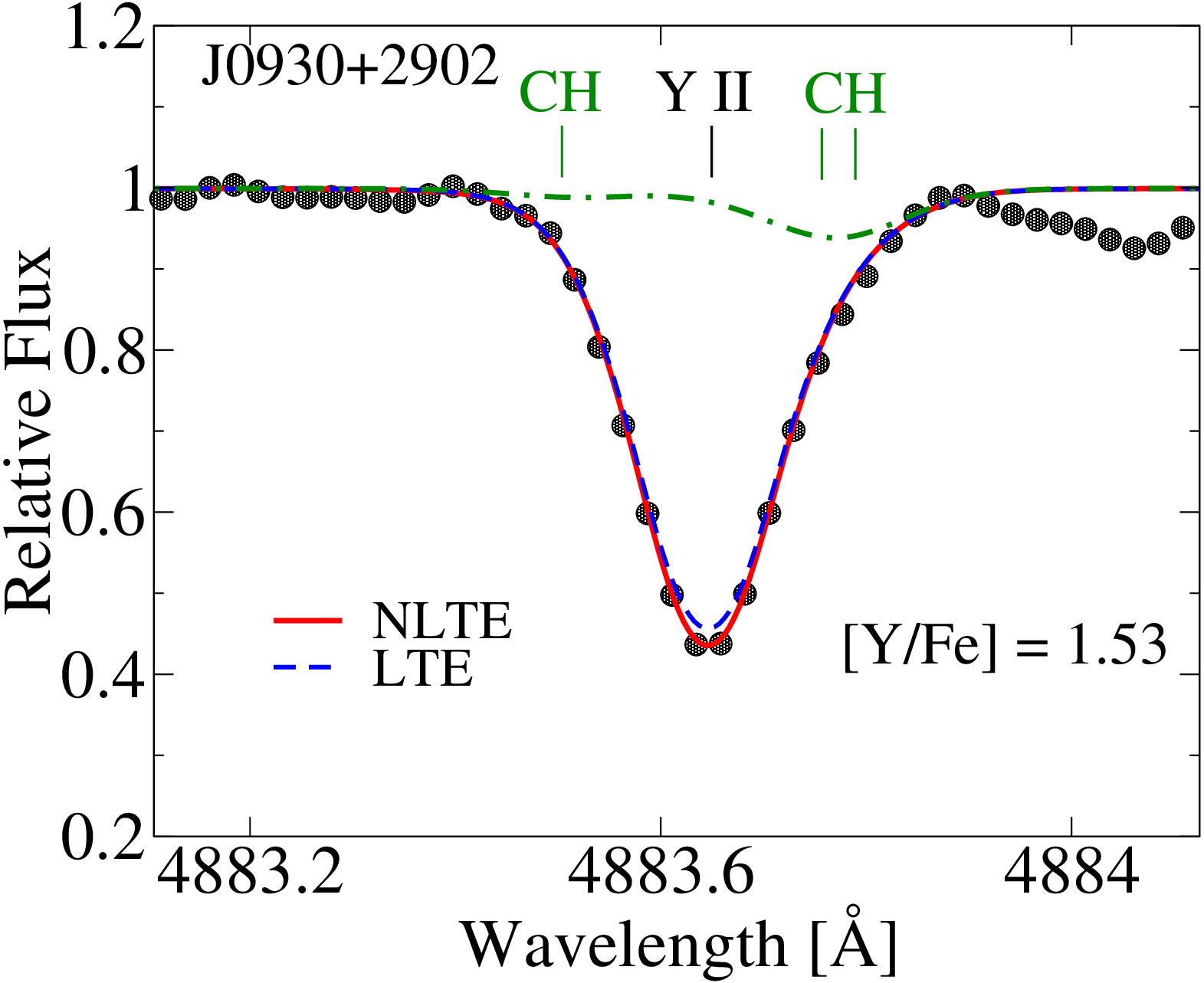}\\
  \centering}
   \parbox{0.32\linewidth}{\includegraphics[scale=0.2]{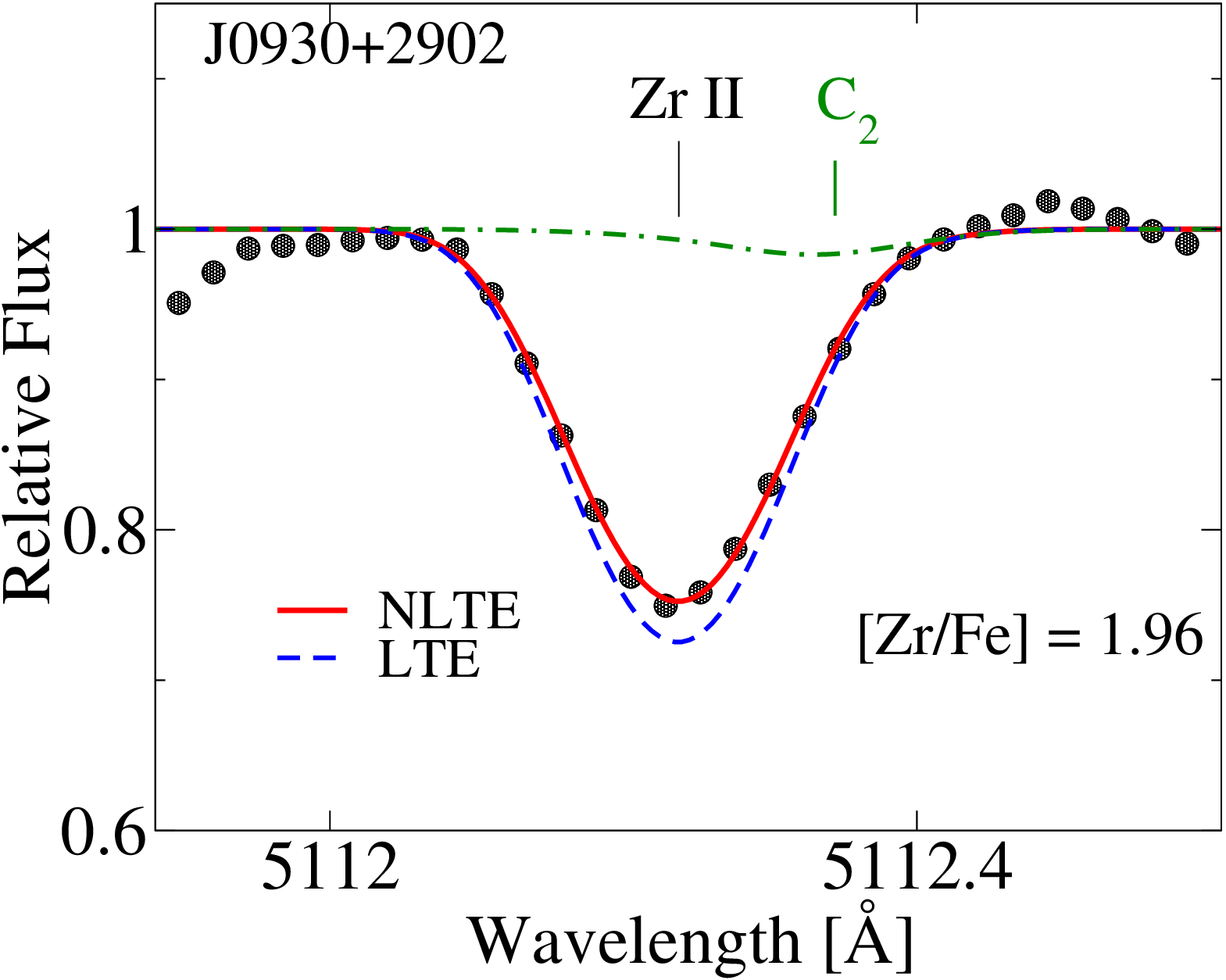}\\
  \centering}
   \parbox{0.32\linewidth}{\includegraphics[scale=0.2]{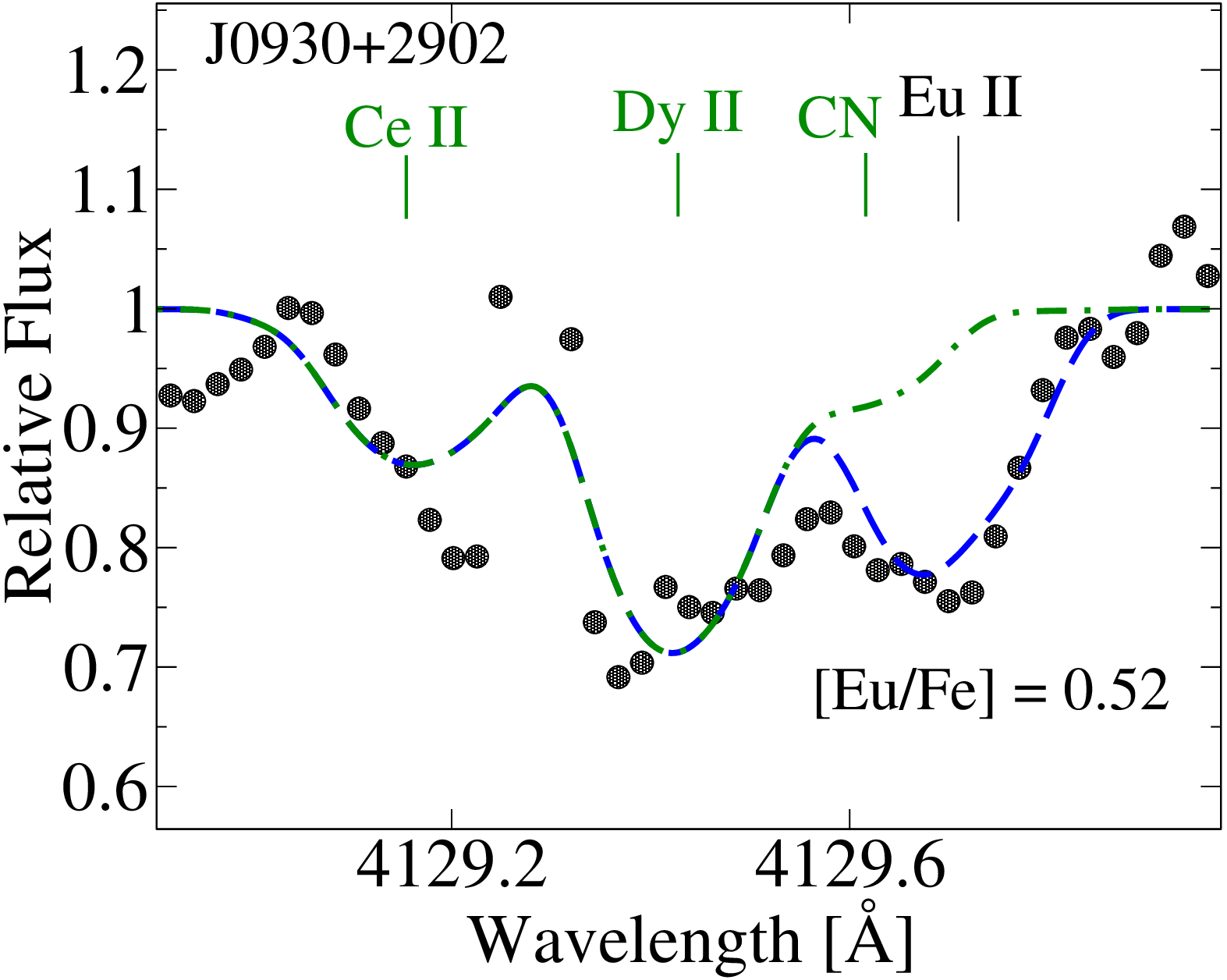}\\
  \centering}
 \hspace{1\linewidth}
 \hfill
 \\[0ex]
 \caption{Best fits of \ion{Y}{2} $\lambda$4883, \ion{Zr}{2} $\lambda$5112, and
\ion{Eu}{2} $\lambda$4129. The observed spectra are shown by bold dots. The symbol sizes
indicate the typical uncertainties of the observed data points. The red solid, blue dashed,
and green dash-dotted curves have the same meanings as in Figure~\ref{Carbon}.}
 \label{YZr}
 \end{center}
 \end{minipage}
 \end{figure*}

Among the heavy neutron-capture elements, the three strong \ion{Ba}{2} lines at $\lambda\lambda$5853,
6141, and 6496~\AA\ yielded an NLTE abundance of [Ba/Fe] = $1.62 \pm 0.14$, with HFS taken into account.
Representative fits to the \ion{Y}{2} $\lambda$4883, \ion{Zr}{2} $\lambda$5112, and \ion{Eu}{2} $\lambda$4129
lines are shown in Figure~\ref{YZr}. We derived [La/Fe] = $1.48 \pm 0.09$ from four \ion{La}{2} lines,
[Ce/Fe] = $1.29 \pm 0.01$ from three \ion{Ce}{2} lines, [Pr/Fe] = 1.23 from only one
\ion{Pr}{2} line, [Nd/Fe] = $1.00 \pm 0.11$ from three \ion{Nd}{2} lines, and
[Sm/Fe] = $0.76 \pm 0.03$ from two \ion{Sm}{2} lines. Spectrum synthesis of the blended
\ion{Eu}{2} $\lambda$4129 feature yielded an NLTE abundance of [Eu/Fe] = 0.68. The strong
enhancements of Ba, La, Ce, and Nd relative to Eu indicate that the neutron-capture element abundance
pattern is dominated by s-process nucleosynthesis, consistent with the abundance characteristics
of CEMP-s stars.

The uncertainty in [X/Fe] has two main sources: line-to-line scatter and uncertainties in the adopted
stellar parameters. The former was taken as the standard deviation of abundances derived from individual
spectral lines, and was listed in Table~\ref{abn2}. The latter was evaluated by varying each parameter
independently by $\Delta T_{\rm eff}=\pm80$~K, $\Delta\log g=\pm0.3$~dex, and
$\Delta\xi_{\rm t}=\pm0.2$~km\,s$^{-1}$ and repeating the abundance analysis. The total uncertainty
was computed by adding these contributions in quadrature (Table~\ref{errors}). For
species represented by a single line, we adopted 0.1 as the scatter contribution to the total uncertainty.
\begin{table}
\centering
\setlength\tabcolsep{1pt}
\caption{Chemical abundances of J0930+2902 from individual lines.}
\label{abn1}
\begin{tabular}{lcccrrr}
\hline
Species     & Wavelength & EP & $\log gf$ & Ref. & A(X)$_{\rm LTE}$ & A(X)$_{\rm NLTE}$ \\
            & \AA\       &  ev  &           &      & dex        & dex   \\\hline
\ion{C}{1}   & 4932.049  & 7.685 & -1.658 & 1 & 7.86    &  7.84 \\
\ion{C}{1}   & 5052.167  & 7.685 & -1.303 & 1 & 7.94    &  7.89 \\
\nodata      & \nodata   & \nodata & \nodata  & \nodata & \nodata & \nodata\\
\ion{O}{1}   & 6300.304  & 0.000 & -9.776 & 5 & $\leq$6.80   & \nodata\\
\ion{Na}{1}  & 5682.633  & 2.102 & -0.706 & 6 & 5.44    &  5.31 \\
\ion{Na}{1}  & 5688.205  & 2.105 & -0.452 & 6 & 5.42    &  5.27 \\
\ion{Na}{1}  & 6154.225  & 2.102 & -1.547 & 6 & 5.52    &  5.43 \\
\ion{Na}{1}  & 6160.747  & 2.105 & -1.246 & 6 & 5.58    &  5.49 \\
\ion{Mg}{1}  & 4571.096  & 0.000 & -5.397 & 7 & 5.40    &  5.75 \\
\ion{Mg}{1}  & 5172.684  & 2.712 & -0.363 & 7 & 5.74    &  5.80 \\
\ion{Mg}{1}  & 5528.405  & 4.346 & -0.547 & 7 & 5.69    &  5.67 \\
\ion{Mg}{1}  & 5711.088  & 4.346 & -1.742 & 7 & 5.74    &  5.74 \\
\nodata      & \nodata   & \nodata & \nodata  & \nodata & \nodata\\
\nodata      & \nodata   & \nodata & \nodata  & \nodata & \nodata\\
\hline
\end{tabular}
\begin{tablenotes}
\scriptsize
\centering
\item References: 
1. \citet{Hibbert1993}; 2. \citet{Brooke2013}; 3. \citet{Masseron2014}; 4. \citet{Brooke2014}; 5. \citet{BalujaZeippen1988};
6. \citet{Karinkuzhi2021}; 7. \citet{Pehlivan2017}; 8. \citet{Garz1973}; 9. \citet{SmithRaggett1981}; 10. \citet{Smith1988};
11. \citet{Aldenius2009}; 12. \citet{Lawler2019}; 13. \citet{Pickering2001}; 14. \citet{Wood2013}; 15. \citet{Ryabchikova1994};
16. \citet{Sobeck2007}; 17. \citet{DenHartog2011}; 18. \citet{Fuhr1988}; 19. \citet{Wickliffe1997}; 20. \citet{OBrian1991};
21. \citet{Bard1994}; 22. \citet{Bard1991}; 23. \citet{RaassenUylings1998}; 24. \citet{Grevesse1999}; 25. \citet{Kurucz2017};
26. \citet{Hannaford1982}; 27. \citet{Ljung2006}; 28. \citet{Cowley1983}; 29. \citet{Gallagher1967}; 30. \citet{Lawler2001a};
31. \citet{Lawler2009}; 32. \citet{Ivarsson2001}; 33. \citet{DenHartog2003}; 34. \citet{Lawler2006}; 35. \citet{Lawler2001b};
36. \citet{Biemont2000}.

\item (The complete table is available in machine-readable format.)
\end{tablenotes}
\end{table}
\begin{table}
\centering
\caption{Abundance ratios [X/Fe] derived for all analyzed species in J0930+2902.}
\label{abn2}
\begin{tabular}{lccccc}
\hline
Species        & $N$ & [X/Fe]$_{\rm LTE}$ & $\sigma$ & [X/Fe]$_{\rm NLTE}$ & $\sigma$ \\
             &  & dex &          &  dex &        \\\hline
$^{12}$C/$^{13}$C        &    & 7.4    &\nodata&\nodata&\nodata \\
\ion{C}{1}  & 2  & 1.87 & 0.06 & 1.84  & 0.04 \\
C (C$_2$)       & 4  & 1.80 & 0.02 &\nodata&\nodata \\
C (CH)          & 7  & 1.80 & 0.02 &\nodata&\nodata \\
N (CN)          & 8  & 2.06 & 0.05 & \nodata& \nodata  \\
\ion{O}{1}      & 1  & $\leq$0.51 & \nodata & \nodata& \nodata  \\
\ion{Na}{1}     & 4  & 1.65 & 0.07 & 1.54  & 0.10  \\
\ion{Mg}{1}     & 4  & 0.44 & 0.16 & 0.54  & 0.05  \\
\ion{Si}{1}     & 1  & 0.45 & \nodata & \nodata& \nodata  \\
\ion{Ca}{1}     & 8  & 0.17 & 0.07 & 0.26 & 0.08 \\
\ion{Sc}{2}     & 4  & 0.18 & 0.08 & \nodata & \nodata \\
\ion{Ti}{2}     & 11 & 0.16 & 0.06 & 0.18 & 0.10 \\
\ion{Cr}{1}     & 5  & -0.17 & 0.23 & 0.30 & 0.24 \\
\ion{Mn}{1}     & 1  & -0.29 & \nodata & 0.09 & \nodata \\
\ion{Co}{1}     & 2  & -0.12 & 0.04 & 0.68 & 0.00 \\
\ion{Ni}{1}     & 2  & -0.11 & 0.10 & \nodata & \nodata \\
\ion{Sr}{2}     & 2  & 1.47 & 0.23 & 1.45  & 0.23  \\
\ion{Y}{2}      & 3  & 1.58 & 0.09 & 1.54  & 0.09  \\
\ion{Zr}{2}     & 3  & 1.92 & 0.01 & 2.02  & 0.00  \\
\ion{Ba}{2}     & 3  & 1.98 & 0.18 & 1.62  & 0.14  \\
\ion{La}{2}     & 4  & 1.48 & 0.09 & \nodata & \nodata  \\
\ion{Ce}{2}     & 3  & 1.29 & 0.01 & \nodata & \nodata  \\
\ion{Pr}{2}     & 1  & 1.23 & \nodata & \nodata & \nodata  \\
\ion{Nd}{2}     & 3  & 1.00 & 0.11 & \nodata & \nodata  \\
\ion{Sm}{2}     & 2  & 0.76 & 0.03 & \nodata & \nodata  \\
\ion{Eu}{2}     & 1  & 0.52 & \nodata & 0.68  & \nodata  \\
\ion{Pb}{1}     & 1  & $\leq$1.74 & \nodata & \nodata &\nodata  \\
\hline
\end{tabular}
\end{table}
\begin{table}
\centering
\caption{Uncertainties in the derived abundances due to stellar parameters.}
\label{errors}
\begin{tabular}{lccccc}
\hline
Ratio    &  $\sigma$  & $\Delta T_{\rm eff}$ & $\Delta \log g$ & $\Delta \xi_{\rm t}$ & $\sigma_{\rm tot}$ \\
         &    & (80~K) & (0.3~dex) & (0.2~km\,s$^{-1}$) &  \\\hline
[\ion{Fe}{2}/H]         & 0.08  &  -0.01  &  0.10   &  -0.01   &  0.13 \\
$[$C/Fe$]$ (\ion{C}{1}) & 0.04  &   0.07  &  0.01   &   0.01   &  0.08 \\
$[$C/Fe$]$ (C$_2$)      & 0.02  &  -0.10  & -0.15   &   0.01   &  0.18 \\
$[$C/Fe$]$ (CH)         & 0.02  &  -0.16  & -0.20   &  -0.02   &  0.26 \\
$[$N/Fe$]$ (CN)         & 0.05  &  -0.22  & -0.19   &   0.01   &  0.29 \\
$[$\ion{Na}{1}/Fe$]$    & 0.10  &  -0.02  & -0.10   &   0.00   &  0.14 \\
$[$\ion{Mg}{1}/Fe$]$    & 0.05  &  -0.06  & -0.10   &  -0.01   &  0.13 \\
$[$\ion{Si}{1}/Fe$]$    & 0.10  &  -0.04  &  0.00   &  -0.01   &  0.11 \\
$[$\ion{Ca}{1}/Fe$]$    & 0.08  &  -0.05  &  0.00   &  -0.02   &  0.10 \\
$[$\ion{Sc}{2}/Fe$]$    & 0.08  &  -0.04  &  0.10   &  -0.04   &  0.14 \\
$[$\ion{Ti}{2}/Fe$]$    & 0.10  &  -0.04  &  0.10   &  -0.04   &  0.15 \\
$[$\ion{Cr}{1}/Fe$]$    & 0.24  &  -0.09  & -0.01   &  -0.06   &  0.26 \\
$[$\ion{Mn}{1}/Fe$]$    & 0.10  &  -0.10  & -0.03   &  -0.09   &  0.17 \\
$[$\ion{Co}{1}/Fe$]$    & 0.00  &  -0.10  &  0.00   &  -0.02   &  0.11 \\
$[$\ion{Ni}{1}/Fe$]$    & 0.10  &  -0.05  &  0.01   &   0.00   &  0.11 \\
$[$\ion{Sr}{2}/Fe$]$    & 0.23  &  -0.08  & -0.10   &   0.00   &  0.26 \\
$[$\ion{Y}{2}/Fe$]$     & 0.09  &  -0.01  &  0.00   &  -0.10   &  0.13 \\
$[$\ion{Zr}{2}/Fe$]$    & 0.00  &  -0.02  &  0.00   &  -0.04   &  0.04 \\
$[$\ion{Ba}{2}/Fe$]$    & 0.14  &  -0.09  & -0.10   &  -0.04   &  0.20 \\
$[$\ion{La}{2}/Fe$]$    & 0.09  &  -0.03  &  0.00   &  -0.08   &  0.12 \\
$[$\ion{Ce}{2}/Fe$]$    & 0.01  &  -0.05  &  0.00   &  -0.01   &  0.05 \\
$[$\ion{Pr}{2}/Fe$]$    & 0.10  &  -0.06  &  0.11   &   0.00   &  0.16 \\
$[$\ion{Nd}{2}/Fe$]$    & 0.11  &  -0.05  &  0.00   &   0.01   &  0.12 \\
$[$\ion{Sm}{2}/Fe$]$    & 0.03  &  -0.02  &  0.11   &   0.00   &  0.12 \\
$[$\ion{Eu}{2}/Fe$]$    & 0.10  &  -0.05  &  0.00   &   0.01   &  0.11 \\\hline
\end{tabular}
\end{table}

\subsection{Abundance Pattern}

\begin{figure*}
 \begin{center}
 \includegraphics[width=14cm]{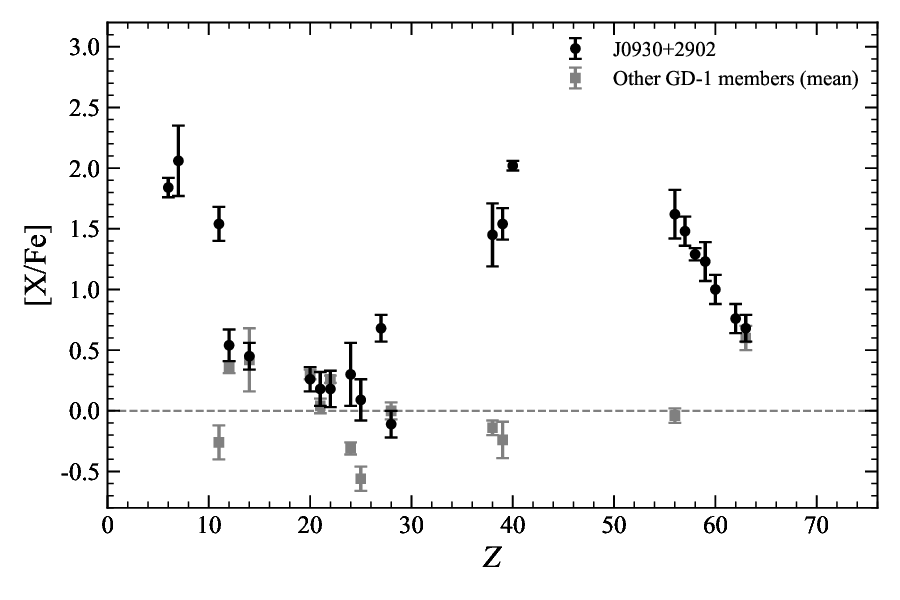}
 \caption{Abundance pattern of J0930+2902 (black circles). For comparison, the mean abundance pattern
 of the six other GD-1 members from Paper~I is shown by gray squares. The error bars on the gray
 squares represent the scatter in the individual abundances of each element among the six other GD-1 members.}
 \label{abun-pattern}
 \end{center}
 \end{figure*}

The chemical abundance pattern of J0930+2902 is characterized by strong enhancements in carbon,
nitrogen, sodium, and neutron-capture elements, as shown in Figure~\ref{abun-pattern}.
For comparison, the mean abundances of the other six GD-1 members analyzed in Paper~I are
also shown. These stars have metallicities similar to that of J0930+2902 and therefore provide
a useful reference for assessing its chemical peculiarities. The abundance ratios of the
$\alpha$-elements Mg, Si, Ca, and Ti, as well as the iron-group elements Sc and Ni, are similar
to those of the other GD-1 members and typical metal-poor halo stars (see Figure~3 of Paper~I).
In contrast, J0930+2902 exhibits a strong enhancement in the odd-$Z$ element sodium. A direct
comparison of carbon and nitrogen abundances with those of the other GD-1 members is not
possible because these elements were not measured in Paper~I.

More striking differences are found among the neutron-capture elements. The other GD-1 members
exhibit nearly solar abundance ratios for Sr, Y, and Ba, whereas J0930+2902 shows strong
enhancements in the s-process elements Sr, Y, Zr, Ba, La, Ce, Pr, and Nd. In contrast, Eu is
only moderately enhanced, at a level similar to that observed in the other GD-1 members.

The broadly similar abundance patterns of the $\alpha$-, iron-group, and r-process elements
suggest that J0930+2902 and the other GD-1 members formed in a comparable chemical environment.
However, the pronounced enhancements of carbon and s-process elements indicate a distinct
enrichment history for J0930+2902. Together with its low metallicity ([Fe/H] $< -1.0$), carbon
enhancement ([C/Fe] $> 1.0$), and s-process enhancement ([Ba/Fe] $> 1.0$ and [Ba/Eu] $> 0.5$),
these abundance characteristics classify J0930+2902 as a CEMP-s star \citep{Beers2005}.

\subsection{Evidence for Binary Mass Transfer}

The abundance pattern of J0930+2902 provides evidence for mass transfer from a former AGB companion.
The carbon isotopic ratio is particularly sensitive to internal mixing and the evolutionary history
of carbon stars \citep{Herwig2005,Abia2017,Ziurys2020}. We derive $^{12}$C/$^{13}$C = 7.4
for J0930+2902. Such a low ratio is commonly observed in evolved CEMP-s stars and, together
with the fact that J0930+2902 is not at an evolutionary stage where it could have self-polluted, suggests
that the observed abundance pattern results from mass transfer from a former AGB companion
\citep{Cohen2006,Lucatello2011,Ram2014}.

J0930+2902 also exhibits substantial enhancements in nitrogen and sodium, with [N/Fe] = 2.06 and
[Na/Fe] = 1.54. The large N enhancement, together with the low $^{12}$C/$^{13}$C ratio, indicates
substantial CN-cycle processing of the transferred material and/or subsequent internal mixing \citep{Herwig2005,Abia2017}.
The strong Na enhancement provides further evidence for material processed in the Ne-Na cycle in
the former AGB companion \citep{Gallino1998,Stancliffe2009,Bisterzo2011}. The coexistence of enhanced
C, N, Na, and s-process elements therefore supports a common origin in AGB nucleosynthesis and strongly
favors a binary mass-transfer scenario for J0930+2902.

\section{Binarity}

The CEMP-s nature of J0930+2902 motivates a search for binarity, for which RV variations
provide one of the most reliable diagnostics. J0930+2902 was first observed by
LAMOST on 2013 March 13, yielding an RV of 115.6~km\,s$^{-1}$ from the LAMOST~DR2 spectrum \citep{Luo2015}.
Approximately seven years later, it was re-observed with Subaru/HDS on 2020 March 11, and an RV of
46.4~km\,s$^{-1}$ was measured \citep{Zhao2025}. The large RV difference of 69.2~km\,s$^{-1}$ between
the two epochs provides strong evidence that J0930+2902 is a binary system, supporting the scenario in which
its chemical peculiarities originated through mass transfer from a former AGB companion.

In addition to RV variations, the presence of a white dwarf (WD) companion can also be
probed through ultraviolet (UV) excess. Because WDs are substantially hotter than cool giant stars,
they can contribute detectable flux at near-ultraviolet (NUV) wavelengths, producing an excess over that
expected from the primary star alone \citep{Vitense2000,Shetye2020,Yang2026}. To search for such a signature,
we modeled the spectral energy distribution (SED) of J0930+2902 and compared the predicted UV flux with
available photometric observations.

The stellar atmosphere corresponding to the adopted atmospheric parameters was generated by
interpolating the ATLAS9 model grid \citep{Castelli2003,Castelli2004} with the IDL package KMOD, which has been
widely used in stellar abundance analyses \citep[e.g.,][]{Meszaros2013,Yong2013}. Synthetic spectra
were then computed with SPECTRUM \citep{Gray1994}. The abundances of the 22 elements derived in this
work were adopted for the spectral synthesis, while the abundances of the remaining elements were assumed
to follow the solar abundance pattern of \citet{Asplund2009}.

\begin{figure}
\centering
\includegraphics[width=14cm]{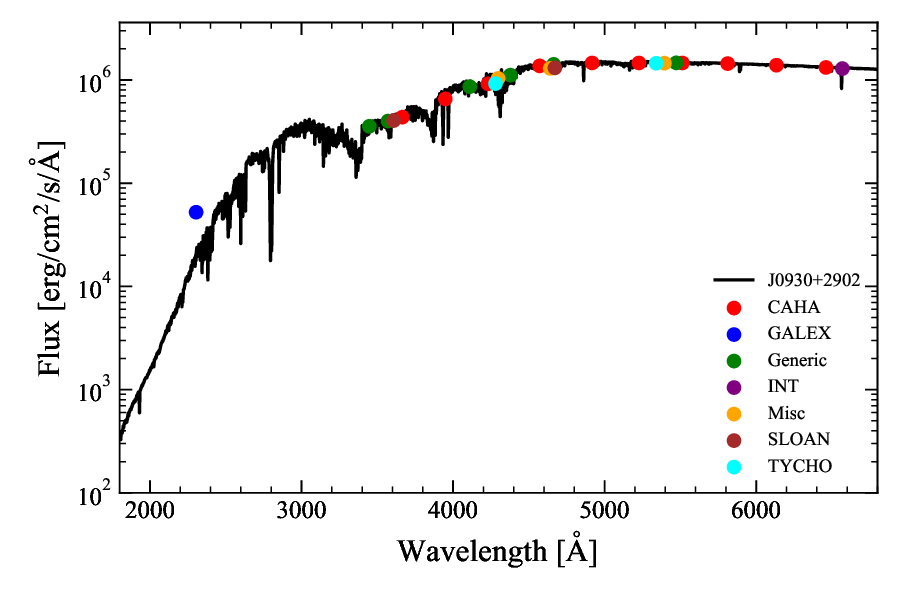}
\caption{SED of J0930+2902. The filled blue circle denotes the GALEX NUV measurement,
while photometric measurements from CAHA, SDSS, and other surveys are shown for comparison. The
solid curve represents the best-fitting synthetic spectrum based on the ATLAS9 model atmosphere.}
\label{fig:vosa}
\end{figure}

Figure~\ref{fig:vosa} compares the best-fitting synthetic SED with multi-band photometry, including
the GALEX NUV measurement \citep{Bianchi2017} together with optical data from CAHA, SDSS, and other
surveys. The synthetic spectrum reproduces the observed optical photometry ($\lambda \gtrsim 3400$~\AA)
well, whereas the observed NUV flux is significantly higher than predicted. This UV excess is naturally
explained by the presence of a hot WD companion. Combined with the large RV variation, the
SED analysis provides independent evidence that J0930+2902 is a binary system, supporting the widely
accepted scenario in which CEMP-s stars are enriched through mass transfer from a former AGB companion
that has since evolved into a WD.

\section{Companion Mass Estimates}

The abundance patterns of neutron-capture elements in CEMP-s stars provide important constraints
on the neutron exposure during s-process nucleosynthesis, as well as on the initial mass and
metallicity of the polluting AGB companion \citep{Busso2001,Bisterzo2012}. We therefore compared
the neutron-capture abundances of J0930+2902 with AGB nucleosynthesis models from the FRUITY database \citep{Cristallo2009,Cristallo2011}, which covers AGB stars with initial masses of 1.3--6.0~$M_{\odot}$
over a wide range of metallicities. These models assume that the primary neutron source is the
$^{13}$C($\alpha$,n)$^{16}$O reaction operating in the radiative $^{13}$C pocket, while the efficiency
of the third dredge-up (TDU) depends on both stellar mass and metallicity. Furthermore, as shown in
Figure~\ref{abun-pattern}, the Eu abundance of J0930+2902, [Eu/Fe] = 0.68, is comparable to those
of the other six GD-1 members, suggesting that its primordial r-process enrichment was likely similar to
that of the other stars in the stream. Assuming that the observed Eu abundance predominantly reflects
this primordial r-process contribution, we estimated the primordial r-process abundance pattern using
the solar-system r-process residuals of \citet{Bisterzo2014} and scaled the pattern to the observed
[Eu/Fe] = 0.68. The scaled primordial r-process contribution was then subtracted from the observed
neutron-capture abundances in linear abundance space:
\begin{equation}
[{\rm X/Fe}]_{\rm res} = \log\left[10^{[{\rm X/Fe}]_{\rm obs}} - 10^{[{\rm X/Fe}]_{\rm r}}\right],
\end{equation}
where [X/Fe]$_{\rm r}$ denotes the scaled primordial r-process contribution and
[X/Fe]$_{\rm res}$ represents the residual neutron-capture abundance after subtracting the
primordial r-process component. This residual abundance pattern was then used to evaluate the
contribution from the former AGB companion. Because the r-process pattern was normalized to
the observed Eu abundance, Eu was not used as an independent constraint in the subsequent
AGB model comparison.

Because the material transferred from the former AGB companion is expected to have been mixed into
the envelope of J0930+2902, the theoretical surface abundances cannot be compared directly with
residual abundances derived above. A dilution factor, $\delta$, is therefore
introduced to account for the mixing of AGB ejecta with the envelope of the accreting star.
Following \citet{Cseh2022}, the diluted abundance of each element was calculated as

\begin{equation}
\rm [X/Fe]_{dil} =
\log \left[(1-\delta)\,10^{[X/Fe]_{ini}}
+\delta\,10^{[X/Fe]_{AGB}}\right],
\end{equation}

where [X/Fe]$_{\rm ini}$ denotes the initial abundance of the star, and [X/Fe]$_{\rm AGB}$ is
the abundance predicted by the FRUITY models.
The resulting diluted abundance, [X/Fe]$_{\rm dil}$, was compared with the residual abundance,
[X/Fe]$_{\rm res}$, obtained after subtracting the primordial r-process contribution from the observed
abundance. Thus, the FRUITY models were used to reproduce the neutron-capture abundance pattern
remaining after removal of the primordial r-process component. Since Ba, La, Ce, and Nd belong to the
same s-process peak, they are expected to exhibit similar abundance enhancements. Following \citet{Cseh2022},
we adopted Ce as the reference element for determining the dilution factor. The dilution
factor was determined by requiring the diluted FRUITY model to reproduce the residual [Ce/Fe]
abundance, and the same dilution factor was then applied to all other elements.

\begin{figure}
\centering
\includegraphics[width=14cm]{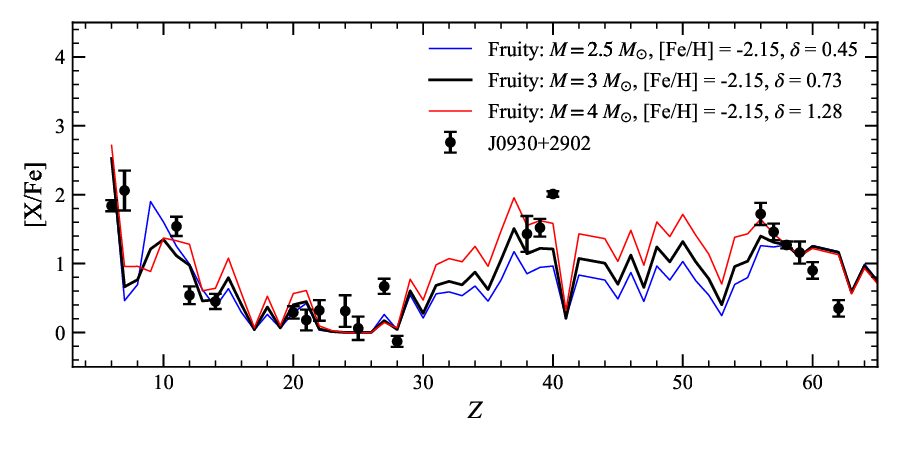}
\caption{Comparison of the elemental abundance pattern of J0930+2902 with FRUITY AGB model
predictions. Filled circles with error bars show the observed abundances for elements from C to Ni, whereas
the abundances of neutron-capture elements from Sr to Sm are the residuals obtained by subtracting the
primordial r-process contribution from the total observed abundances. The neutron-capture residuals are
used to fit the FRUITY AGB models, while the abundances from C to Ni are not included in the fitting and
are shown for comparison with the best-fitting model. The blue, black, and red curves show the 2.5, 3.0,
and 4.0~$M_{\odot}$ models, respectively, with the best-fitting model having $\delta = 0.73$.}
\label{fig:fruity}
\end{figure}

Model comparisons were restricted to metallicities consistent with that of J0930+2902, and the preferred
solution was identified by minimizing the $\chi^2$ difference between the diluted model abundance pattern
and the observed residual abundances of neutron-capture elements. Only models with $\delta < 1$ were
considered, since $\delta$ = 1 would imply an undiluted AGB contribution without any mixing with the original
stellar material. Figure~\ref{fig:fruity} compares the residual neutron-capture abundances of J0930+2902
with the best-fitting diluted FRUITY model with M = 3.0~$M_{\odot}$ and [Fe/H] = $-2.15$. The predicted
abundance pattern provides a good overall match to the residual abundances, with a dilution factor of
$\delta$ = 0.73. To illustrate the sensitivity of the predicted abundances to the AGB mass, models with
M = 2.5 and 4.0~$M_{\odot}$ are also shown. The 2.5~$M_{\odot}$ model provides a poorer match, whereas the
4.0~$M_{\odot}$ model requires $\delta > 1$ and is therefore excluded. These results suggest that the
chemical peculiarities of J0930+2902 can be explained by mass transfer from an AGB companion with an
initial mass of approximately 3.0~$M_{\odot}$. This inferred mass is consistent with previous studies
suggesting that the polluting companions of CEMP-s stars are predominantly low-mass AGB stars
\citep[e.g.,][]{Jorissen2019,Roriz2021,Yang2026}. For comparison, the observed abundances of
the light, $\alpha$-, and iron-group elements are also shown in Figure~\ref{fig:fruity}. The model
predicts enhanced C, N, and Na, although it does not reproduce the observed abundances well. We note
that the inferred AGB abundance pattern depends on the adopted r-process decomposition. This issue
can be further investigated in future work with a more detailed neutron-capture abundance pattern.

\section{Implications for the GD-1 Progenitor}

The presence of a CEMP-s star in the GD-1 stream is particularly intriguing because
stars showing signatures of binary AGB mass transfer are extremely rare in present-day GCs.
Although CH and Ba stars have been identified in Galactic GCs \citep{Cote1997,DOrazi2010,DaCosta2025},
they constitute only a very small fraction of the cluster populations. For J0930+2902, the strong C and
s-process enhancements, together with the large RV variation and NUV excess, provide independent
evidence for such a mass-transfer system.
Our abundance analysis favors an AGB donor with an initial mass of approximately
$3~M_\odot$. Such a star evolves to the thermally pulsing AGB (TP-AGB) phase on a timescale of order
$10^8$~yr, during which C- and s-process material can be produced and transferred to a lower-mass
companion \citep{Cristallo2011,Cristallo2015}. For the GD-1 stream, N-body modeling suggests that
the present-day stream has been forming over approximately 2--3~Gyr \citep{Webb2019}.
The substantially shorter evolutionary timescale of the donor leaves a plausible time window
for the binary to be tidally stripped from the progenitor cluster before the donor reached the
TP-AGB phase. If such stripping occurred, J0930+2902 could have entered the
lower-density stream environment before mass transfer took place, allowing the binary to survive
until the donor reached the TP-AGB phase. We therefore consider tidal stripping of
the binary before the onset of the TP-AGB phase, followed by mass transfer in the
lower-density stream environment, to be a plausible evolutionary scenario.

However, these timescales do not uniquely determine the relative timing of Galactic
accretion, tidal disruption, and mass transfer. The time of complete dissolution of the progenitor
cluster remains uncertain, with scenarios involving final dissolution $\sim$0.5~Gyr ago or more than
2.5~Gyr ago \citep{Webb2019}. Alternatively, the binary may have remained associated with the
progenitor cluster until the donor reached the TP-AGB phase and underwent mass transfer.
The available observations do not distinguish between these possibilities. Nevertheless, the presence
of J0930+2902 demonstrates that the chemical signature produced by AGB binary mass transfer
survived the subsequent tidal evolution and dissolution of the progenitor cluster and remains
detectable in the GD-1 stream.

\section{Conclusions}

We report the discovery of J0930+2902, the first CEMP-s star identified in the GD-1 stellar stream.
Using high-resolution Subaru/HDS spectroscopy, we performed a comprehensive chemical abundance analysis,
investigated its binary nature, and constrained the properties of its former AGB companion through
comparisons with theoretical nucleosynthesis models. Our main conclusions are summarized as follows.

\begin{enumerate}

\item We derived the stellar atmospheric parameters of J0930+2902 using Gaia photometry and parallaxes,
together with an iterative NLTE analysis of Fe lines. Chemical abundances were determined for 22 elements,
including NLTE analyses of C, Na, Mg, Ca, Ti, Cr, Mn, Co, Sr, Y, Zr, Ba, and Eu. J0930+2902 exhibits
pronounced enhancements of C, N, Na, and s-process elements, with [C/Fe] = 1.84, [Ba/Fe] = 1.62,
and [Ba/Eu] $= 0.94$. Together with its low carbon isotopic ratio, $^{12}$C/$^{13}$C = 7.4, these abundance
characteristics firmly classify J0930+2902 as a CEMP-s star and indicate enrichment through mass transfer
from a former AGB companion.

\item J0930+2902 exhibits [X/Fe] ratios of $\alpha$- and iron-group elements similar to those of the
other GD-1 members, whereas its strong enhancements of C, N, Na, and s-process elements clearly
distinguish it from the rest of the stream. This indicates that, although J0930+2902 likely formed
in a chemical environment similar to that of the other GD-1 members, its surface composition was subsequently
modified by binary mass transfer rather than by the global chemical evolution of the GD-1 progenitor. The
preservation of these chemical signatures demonstrates that binary mass transfer can leave
long-lived abundance imprints in an ancient disrupted stellar system.

\item Independent evidence for binarity is provided by the large RV variation between
the LAMOST and Subaru observations and the significant NUV excess revealed by the
spectral energy distribution, both of which indicate the presence of a WD companion.
Comparison of the residual neutron-capture element abundance pattern, obtained by subtracting the
estimated r-process contribution from the observed abundances, with FRUITY AGB nucleosynthesis models
suggests that this residual pattern is best reproduced by mass transfer from an AGB star
with an initial mass of approximately 3.0~$M_{\odot}$.

\end{enumerate}

J0930+2902 provides a valuable probe of binary evolution in the GD-1 progenitor.
The observed mass-transfer signature implies that the binary remained sufficiently intact
for the $\sim3~M_\odot$ donor to evolve to the TP-AGB phase, which occurs on
a timescale of several $10^8$~yr. This may be consistent with a progenitor environment in which
the binary avoided significant dynamical disruption, for example because of a relatively low
stellar density or a sufficiently hard binary. Alternatively, the binary may have been tidally
stripped from the progenitor before the donor reached the AGB phase and subsequently undergone
mass transfer in the lower-density stream environment. The relative timing of the mass-transfer event,
tidal disruption, and accretion of the GD-1 progenitor by the Milky Way, however, remains uncertain.
A larger, uniformly selected sample is needed to determine whether such systems are unusually common
in the GD-1 progenitor or are consistent with the expected rare occurrence of surviving mass-transfer
binaries.

Our results demonstrate that the chemical signature of binary mass transfer can survive the
subsequent tidal evolution and disruption of the parent stellar system, providing a new perspective
on the origin and chemical evolution of ancient accreted stellar populations. Future RV
monitoring and an orbital solution will provide stronger constraints on the
properties of the WD companion and further test the proposed binary mass-transfer scenario.

\section*{Data Availability}

The data underlying this paper are available in machine-readable form.

\section*{Code Availability}
We use the standard data analysis tools as follows: SynthV\_NLTE \citep{Tsymbal2019},
BINMAG \citep{Kochukhov2010,Kochukhov2018}, DETAIL \citep{Mashonkina2011},
iSpec \citep{BlancoCuaresma2014}, MARCS \citep{Gustafsson2008}, TOPCAT \citep{Taylor2005},
Matplotlib \citep{Hunter2007}, NumPy \citep{Harris2020}, SciPy \citep{Virtanen2020},
Jupyter \citep{Kluyver2016}, Astropy \citep{Astropy2013,Astropy2018,Astropy2022},
and Gala \citep{Price2017}.

\begin{acknowledgments}

We thank the referee for the constructive comments and suggestions, which have significantly
improved the manuscript.
This work is supported by the National Natural Science Foundation of China (NSFC) under
grant Nos. 12588202, 12273055, 12503024, and W2432009; the National Key R\&D Program of China under grant
Nos. 2023YFE0107800 and 2024YFA1611902; the International Partnership Program of the
Chinese Academy of Sciences under grant No. 178GJHZ2022040GC; the Independent Research
Program of the National Astronomical Observatories, Chinese Academy of Sciences under grant
No. E4ZB0301; the Strategic Priority Research Program of the Chinese Academy of
Sciences under grant No. XDB1160301; and the Fundamental Research Funds of China
West Normal University under grant Nos. 22kE041 and 2026kc0010. We acknowledge the support from
the 2~m Chinese Space Station Telescope project. This work has been supported by a
Grant-in-Aid for Scientific Research (KAKENHI) (JP22K03688, JP25K01046, JP25HP8011) from
the Japan Society for the Promotion of Science. This research is based on data collected at
Subaru Telescope, which is operated by the National Astronomical Observatory of Japan.
We are honored and grateful for the opportunity of observing the Universe from Maunakea,
which has the cultural, historical, and natural significance in Hawaii. Guoshoujing Telescope
(the Large Sky Area Multi-Object Fiber Spectroscopic Telescope, LAMOST) is a National Major
Scientific Project built by the Chinese Academy of Sciences. Funding for the project has
been provided by the National Development and Reform Commission. It is operated and
managed by the National Astronomical Observatories, Chinese Academy of Sciences.
This work has made use of data from the European Space Agency (ESA) mission Gaia
(https://www.cosmos.esa.int/gaia), processed by the Gaia Data Processing and Analysis Consortium
(DPAC, https://www.cosmos.esa.int/web/gaia/dpac/consortium). Funding for the DPAC has been
provided by national institutions, in particular the institutions participating in the
Gaia Multilateral Agreement. The work was carried out under the state assignment of the
Institute of astronomy of RAS.
\end{acknowledgments}

\end{document}